\documentclass[aps,prd,twocolumn,nofootinbib]{revtex4}

\usepackage{graphicx}
\usepackage{amssymb}
\usepackage{color}
\usepackage{CJK}
\usepackage[colorlinks=true,linkcolor=blue,citecolor=blue, urlcolor=blue]{hyperref}

\begin{document}

\title{Generalized Crewther relation and a novel demonstration of the scheme independence of commensurate scale relations up to all orders}

\author{Xu-Dong Huang$^1$}
\email{hxud@cqu.edu.cn}

\author{Xing-Gang Wu$^{1,2}$}
\email{wuxg@cqu.edu.cn (Corresponding author)}

\author{Qing Yu$^1$}
\email{yuq@cqu.edu.cn}

\author{Xu-Chang Zheng$^1$}
\email{zhengxc@cqu.edu.cn}

\author{Jun Zeng$^3$}
\email{zengj@cqu.edu.cn}

\author{Jian-Ming Shen$^4$}
\email{shenjm@hnu.edu.cn}

\affiliation{$^1$ Department of Physics, Chongqing University, Chongqing 401331, People's Republic of China}
\affiliation{$^2$ Chongqing Key Laboratory for Strongly Coupled Physics, Chongqing 401331, People's Republic of China}
\affiliation{$^3$ INPAC, Key Laboratory for Particle Astrophysics and Cosmology (MOE), Shanghai Key Laboratory for Particle Physics and Cosmology, School of Physics and Astronomy, Shanghai Jiao Tong University, Shanghai 200240, People's Republic of China}
\affiliation{$^4$ School of Physics and Electronics, Hunan University, Changsha 410082, People's Republic of China}

\date{\today}

\begin{abstract}

In the paper, we make a detailed study on the generalized Crewther Relation (GCR) between the Adler function ($D$) and the Gross-Llewellyn Smith sum rules coefficient ($C^{\rm GLS}$) by using the newly suggested single-scale approach of the principle of maximum conformality (PMC). The resultant GCR is scheme-independent, whose residual scale dependence due to unknown higher-order terms are highly suppressed. Thus a precise test of QCD theory without renormalization scheme and scale ambiguities can be achieved by comparing with the data. Moreover, a demonstration of the scheme independence of commensurate scale relation up to all orders has been presented. And as the first time, the Pade approximation approach has been adopted for estimating the unknown $5_{\rm th}$-loop contributions from the known four-loop perturbative series.

\end{abstract}

\maketitle

\section{Introduction}

The Crewther relation~\cite{Crewther:1972kn, Adler:1973kz} provides a non-trivial relation for three fundamental constants, $3S = K R'$, where $S$ is the anomalous constant of $\pi^0\to\gamma\gamma$, $K$ is coefficient of the Bjorken sum rules for the polarized deep-inelastic electron scattering~\cite{Bjorken:1966jh}, and $R'$ is the isovector part of the cross-section ratio for the electron-positron annihilation into hadrons~\cite{Adler:1974gd}. In the theory of quantum chromodynamics (QCD)~\cite{Politzer:1973fx, Gross:1973id}, the Crewther relation is improved as the ``Generalized Crewther Relation (GCR)"~\cite{Bjorken:1967px, Bjorken:1969mm, Broadhurst:1993ru, Brodsky:1995tb, Crewther:1997ux, Gabadadze:1995ei, Braun:2003rp}:
\begin{eqnarray}
D^{\rm NS}(a_s) C^{\rm Bjp}(a_s) &=& 1+\Delta^{*}_{\rm csb}, \label{GCR1}
\end{eqnarray}
or
\begin{eqnarray}
D(a_s) C^{\rm GLS}(a_s) &=& 1+\Delta_{\rm csb},  \label{eq:GCR}
\end{eqnarray}
where $a_s = \alpha_s/\pi$, $D^{\rm NS}$ is the non-singlet Adler function, $C^{\rm Bjp}$ is derived by the Bjorken sum rules for polarized deep-inelastic electron scattering, $D$ is the Adler function, and $C^{\rm GLS}$ is the coefficient of the Gross-Llewellyn Smith (GLS) sum rules~\cite{Gross:1969jf}. The $\Delta^{*}_{\rm csb}$ and $\Delta_{\rm csb}$ are conformal breaking terms, and for example, the $\Delta_{\rm csb}$-term takes the form
\begin{eqnarray}
\Delta_{\rm csb} = \frac{\beta(a_s)}{a_s}K(a_s)=-\sum_{i\geq2}\sum_{k=1}^{i-1} K_k \beta_{i-1-k} a_s^i , \label{eq:csbbeta}
\end{eqnarray}
where $\beta(a_s)=-\sum_{i\geq0}\beta_{i} a_s^{i+2}$ is the usual $\beta$-function, and the coefficients $K_k$ are free of $\{\beta_i\}$-functions.

The perturbative QCD (pQCD) corrections to $D^{\rm NS}$ and the Bjorken sum rules have been computed up to ${\cal O}(\alpha_s^3)$-level~\cite{Gorishnii:1990vf, Surguladze:1990tg, Chetyrkin:1994js, Chetyrkin:1996ez, Larin:1991tj} and ${\cal O}(\alpha_s^4)$-level~\cite{Baikov:2008jh, Baikov:2010je}, respectively. The GCR (\ref{GCR1}) between the non-singlet Adler function and the Bjorken sum rules has been discussed in Refs.\cite{Cvetic:2016rot, Garkusha:2018mua, Shen:2016dnq}. The GCR (\ref{eq:GCR}) between the Adler function and the GLS sum rules has been investigated up to ${\cal O}(\alpha_s^3)$-level in Ref.\cite{Brodsky:1995tb}. Using the known $\alpha_s^4$-order corrections~\cite{Baikov:2010iw, Baikov:2012zn}, we has the chance to derive a more accurate GCR (\ref{eq:GCR}) up to ${\cal O}(\alpha_s^4)$-level. It is well-known that a physical observable is independent to any choice of theoretical conventions such as the renormalization scheme and renormalization scale. This property is called as the ``renormalization group invariance" (RGI)~\cite{Petermann:1953wpa, GellMann:1954fq, Peterman:1978tb, Callan:1970yg, Symanzik:1970rt}. For a fixed-order pQCD prediction, if the perturbative coefficient before $\alpha_s$ and the corresponding $\alpha_s$-value at each order do not well match with each other, then the RGI shall be explicitly broken~\cite{Wu:2013ei, Wu:2014iba}. Conventionally, people adopts the ``guessed" typical momentum flow of the process as the optimal renormalization scale with the purpose of eliminating the large logs to improve the pQCD convergence or minimizing the contributions from the higher-order loop diagrams or achieving theoretical prediction in agreement with the experimental data. Such kind of treatment directly breaks the RGI and reduces the predictive power of pQCD. Thus it is important to have a proper scale-setting approach to achieve a scale-invariant fixed-order prediction.

In the literature, the principle of maximum conformality (PMC)~\cite{Brodsky:2011ta, Brodsky:2012rj, Mojaza:2012mf, Brodsky:2013vpa} has been proposed to eliminate those two artificial ambiguities. The purpose of PMC is not to find an optimal renormalization scale but to fix an effective $\alpha_s$ of the process by using the renormalization group equation (RGE); And the PMC prediction satisfies all self-consistency conditions of the renormalization group~\cite{Brodsky:2012ms}. Two multi-scale approaches have been suggested to achieve the goal of PMC, which are equivalent in sense of perturbative theory~\cite{Bi:2015wea}, and a collection of their successful applications can be found in Ref.\cite{Wu:2015rga}. For the multi-scale approach, the PMC sets the scales via an order-by-order manner; the individual scales at each order reflect the varying virtuality of the amplitudes at those orders. It has been noted that the PMC multi-scale approach has two types of residual scale dependence because of unknown perturbative terms~\cite{Zheng:2013uja}. Those residual scale dependence suffer from both the $\alpha_s$-power suppression and the exponential suppression, but their magnitudes could be large due to poor convergence of the perturbative series of either the PMC scale or the pQCD approximant~\cite{Wu:2019mky}.

Lately, the PMC single-scale approach~\cite{Shen:2017pdu} has been suggested to suppress the residual scale dependence by introducing an overall effective $\alpha_s$. The argument of such effective $\alpha_s$ corresponds to the overall effective momentum flow of the process, which is also independent to any choice of renormalization scale. It has been shown that by using the PMC single-scale approach and the $C$-scheme strong coupling constant~\cite{Boito:2016pwf}, one can achieve a strict demonstration of the scheme-invariant and scale-invariant PMC prediction up to any fixed order~\cite{Wu:2018cmb}. Moreover, the resulting renormalization scheme- and scale- independent conformal series is helpful not only for achieving precise pQCD predictions but also for a reliable prediction of the contributions of unknown higher-orders; some of its applications have been performed in Refs.\cite{Du:2018dma, Yu:2018hgw, Yu:2019mce, Huang:2020rtx, Yu:2020tri}, which are estimated by using the $\rm Pad\acute{e}$ resummation approach~\cite{Basdevant:1972fe, Samuel:1992qg, Samuel:1995jc}. In the present paper, we shall adopt the PMC single-scale approach to deal with the GCR (\ref{eq:GCR}), and then, as the first time, we shall estimate the unknown $5_{\rm th}$-loop contributions for GCR (\ref{eq:GCR}). A novel demonstration of the scheme independence of commensurate scale relation up to all orders shall also be presented.

\section{Generalized Crewther relation under the PMC single-scale approach} \label{II}

It is helpful to define the effective charge for a physical observable~\cite{Grunberg:1980ja, Grunberg:1982fw, Dhar:1983py}, which incorporates the entire radiative correction into its definition. For example, the GLS sum rules indicates that the isospin singlet structure function $xF_3(x,Q^2)$ satisfies an unsubtracted dispersion relation~\cite{Gross:1969jf}. i.e.,
\begin{eqnarray}
\frac{1}{2} \int_0^1 \frac{d x}{x} xF_3(x,Q^2) = 3 C^{\rm GLS}(a_s),
\end{eqnarray}
where $xF_3(x,Q^2)=xF^{\nu p}_3(x,Q^2)+xF^{\bar{\nu} p}_3(x,Q^2)$, and $Q$ is the momentum transfer. The entire radiative QCD corrections can be defined as an effective charge $a_{F_3}(Q)$. Moreover, the Adler function of the cross-section ratio for the electron-positron annihilation into hadrons~\cite{Adler:1974gd}
\begin{eqnarray}
D(Q^2) &=& -12 \pi^2 Q^2 \frac{d}{d Q^2}\Pi(Q^2),  \label{AdlerF}
\end{eqnarray}
where
\begin{eqnarray}
\Pi(Q^2)&=& -\frac{Q^2}{12\pi^2}\int^{\infty}_{4 m_{\pi}^2}\frac{R_{e^+e^-}(s)ds}{s(s+Q^2)}.
\end{eqnarray}
The Adler function $D$ can be defined as the effective charge $a_D(Q)$. Thus the Adler function $D$ and the GLS sum rules coefficient $C^{\rm GLS}$ can be rewritten as
\begin{eqnarray}
D(a_s)&=&1+ a_D(Q),\label{D}\\
C^{\rm GLS}(a_s)&=&1- a_{F_3}(Q).\label{CGLS}
\end{eqnarray}
The effective charges $a_D(Q)$ and $a_{F_3}(Q)$ are by definition pQCD calculable, which can be expressed as the following perturbative form,
\begin{eqnarray}
a_{\cal S}&=&r^{\cal S}_{1,0}a_s + (r^{\cal S}_{2,0}+\beta_{0}r^{\cal S}_{2,1}) a_s^{2}\nonumber\\
&&+(r^{\cal S}_{3,0}+\beta_{1}r^{\cal S}_{2,1}+ 2\beta_{0}r^{\cal S}_{3,1}+ \beta_{0}^{2}r^{\cal S}_{3,2})a_s^{3}\nonumber\\
&& +(r^{\cal S}_{4,0}+\beta_{2}r^{\cal S}_{2,1}+ 2\beta_{1}r^{\cal S}_{3,1} + \frac{5}{2}\beta_{1}\beta_{0}r^{\cal S}_{3,2} \nonumber\\
&&+3\beta_{0}r^{\cal S}_{4,1}+3\beta_{0}^{2}r^{\cal S}_{4,2} +\beta_{0}^{3}r^{\cal S}_{4,3}) a_s^{4}+\mathcal{O}(a^5_s), \label{cDij}
\end{eqnarray}
where ${\cal S}=D$ or ${F_3}$, respectively. $r^{\cal S}_{i,j=0}$ are conformal coefficients with $r^{\cal S}_{1,0}=1$, and $r^{\cal S}_{i,j\neq0}$ are nonconformal coefficients. The $\beta$-pattern at each order is determined by RGE~\cite{Mojaza:2012mf, Brodsky:2013vpa}. The coefficients $r^{D, F_3}_{i,j}$ up to four-loop level under $\overline{\rm MS}$ scheme can be read from Refs.\cite{Baikov:2012zm, Baikov:2010je} by using the general QCD degeneracy relations~\cite{Bi:2015wea}.

\begin{figure}[htb]
\includegraphics[width=0.50\textwidth]{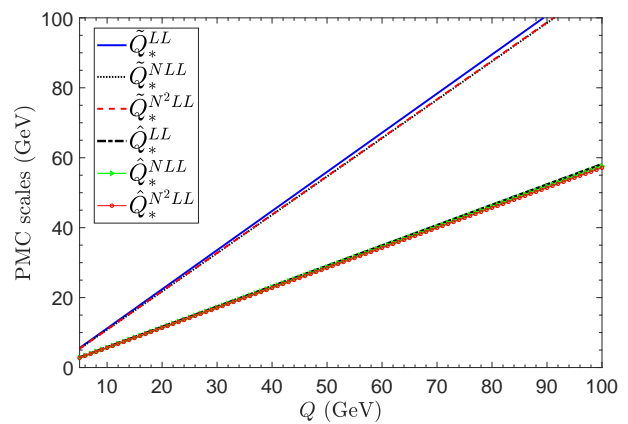}
\caption{The PMC scales ${\tilde Q}_*$ and ${\hat Q}_*$ versus the momentum scale $Q$ up to LL, NLL and N$^2$LL accuracies, respectively.}  \label{Qstar}
\end{figure}

After applying the PMC single-scale approach~\cite{Shen:2017pdu}, all $\{\beta_i\}$-terms are used to fix the correct $\alpha_s$-value by using the RGE, and up to four-loop accuracy, we obtain the following conformal series
\begin{eqnarray}
D(a_s)|_{\rm PMC}&=&1+\left[r^D_{1,0}a_s({\tilde Q}_*)+r^D_{2,0}a^2_s({\tilde Q}_*) \right. \nonumber \\
&& \quad\quad \left. +r^D_{3,0}a^3_s({\tilde Q}_*) +r^D_{4,0}a^4_s({\tilde Q}_*) \right], \\
C^{\rm GLS}(a_s)|_{\rm PMC}&=&1- \left[r^{F_3}_{1,0}a_s({\hat Q}_*) +r^{F_3}_{2,0}a^2_s({\hat Q}_*) \right. \nonumber \\
&& \quad\quad \left. +r^{F_3}_{3,0}a^3_s({\hat Q}_*) +r^{F_3}_{4,0}a^4_s({\hat Q}_*) \right], \label{CGLSpmc}
\end{eqnarray}
where ${\tilde Q}_*$ and ${\hat Q}_*$ are in perturbative series which can be derived from the pQCD series of $a_D$ and $a_{F_3}$. ${\tilde Q}_*$ and ${\hat Q}_*$ correspond to the overall momentum flows of the effective charges $a_D(Q)$ and $a_{F_3}(Q)$, which are independent to any choice of renormalization scale. This property confirms that the PMC is not to choose an optimal renormalization scale, but to find the correct momentum flow of the process. Using the four-loop pQCD series of $a_D$ and $a_{F_3}$, we determine their magnitudes up to N$^2$LL accuracy by replacing the coefficients ${\hat r}_{i,j}$ in the Eqs.~(8-11) of Ref.\cite{Shen:2017pdu} with $r^D_{i,j}$ or $r^{F_3}_{i,j}$, respectively. We present ${\tilde Q}_*$ and ${\hat Q}_*$ up to different accuracies in Fig.~\ref{Qstar}. As shown by Fig.~\ref{Qstar}, because the perturbative series of $a_D$ and $a_{F_3}$ have good perturbative convergence, it is interesting to find that their magnitudes up to different accuracies, such as the leading log (LL), the next-to-leading log (NLL), and the next-next-leading log (N$^2$LL) accuracies, are very close to each other. Thus one can treat the N$^2$LL-accurate ${\tilde Q}_*$ and ${\hat Q}_*$ as their exact values.

\begin{figure}[htb]
\includegraphics[width=0.50\textwidth]{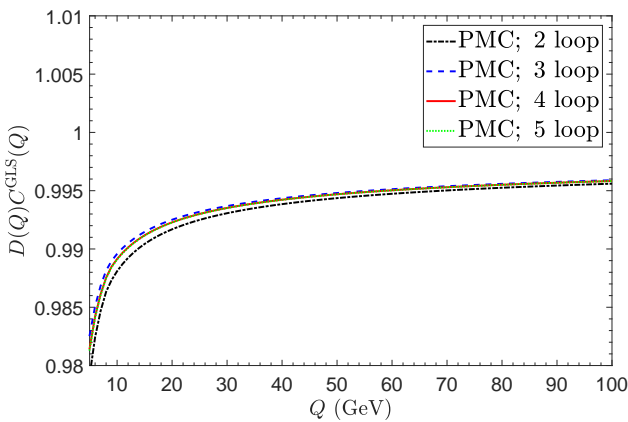}
\caption{The $D(a_s) C^{\rm GLS}(a_s)$ versus the momentum $Q$ under the PMC single-scale approach up to two-loop, three-loop, and four-loop levels, respectively. The uncalculated five-loop result is predicted by using the Pad\'{e} approximation approach. }  \label{RPMC}
\end{figure}

By applying the PMC single-scale approach, one can improve the precision of $D(a_s)$ and $C^{\rm Bjp}(a_s)$. Fig.~\ref{RPMC} shows the PMC predictions of $D(a_s) C^{\rm GLS}(a_s)$ up to two-loop, three-loop, and four-loop levels, respectively. Moreover, the PMC scheme-invariant and scale-invariant conformal series provides a reliable platform for predicting the uncalculated high-order terms~\cite{Du:2018dma}. As the fist time, we also present the uncalculated five-loop result in Fig.~\ref{RPMC}, which is predicted by using the Pad\'{e} approximation approach (PAA)~\cite{Basdevant:1972fe, Samuel:1992qg, Samuel:1995jc} and by using the preferable [N/M]=[0/n-1]-type Pad\'{e} generating function which makes the PAA geometric series be self-consistent with the PMC prediction~\cite{Du:2018dma}. More explicitly, the PAA has been introduced for estimating the $(n+1)_{\rm th}$-order coefficient in a given $n_{\rm th}$-order perturbative series and feasible conjectures on the likely high-order behavior of the series. For example, for the following conformal series
\begin{equation}
\rho(Q)=\sum^n_{i=1} r_{i,0}\; a_s^i,
\end{equation}
for the present cases, we have $n=4$ and $\rho(Q)=a_D|_{\rm PMC}$ or $a_{F_3}|_{\rm PMC}$; its $[N/M]$-type fractional generating function by using the PAA is defined as
\begin{equation}
\rho^{N/M}(Q) = a_s\times\frac{b_0+b_1 a_s+\cdots+b_N a_s^N}{1+c_1 a_s+\cdots+c_M a_s^M},  \label{Paam}
\end{equation}
where $M\geq1$ and $N+M=n-1$. Then the unknown $(n+1)_{\rm th}$-order coefficient $r_{n+1,0}$ can be predicted by using the known coefficients $\{r_{1,0},\cdots,r_{n,0}\}$ via expanding the fractional generating function over $a_s$. That is Eq.(\ref{Paam}) can be reexpressed as
\begin{equation}
\rho^{N/M}(Q) = \sum^n_{i=1} r_{i,0} a_s^i + r_{n+1,0} a_s^{n+1}+\cdots.
\end{equation}
We can first express all the coefficients $\{b_0,\cdots,b_N\}$ and $\{c_1,\cdots,c_M\}$ by the known coefficients $r_{\{1,\cdots,n\},0}$, and then get the coefficient $r_{n+1,0}$ over $\{b_i\}$ and $\{c_i\}$, which can be finally expressed by $\{r_{1,0},\cdots,r_{n,0}\}$. We put the PAA predictions of $a_D|_{\rm PMC}$ and $a_{F_3}|_{\rm PMC}$ in Table.~\ref{pmcpaa1} and Table.~\ref{pmcpaa2}, respectively, in which the known coefficients (``Exact") at different orders are also presented as comparison. They show that when we have known more perturbative terms, the PAA predicted coefficients shall become closer to the ``Exact" ones.

\begin{table}[htb]
\begin{tabular}{ccc}
\hline
 ~~~ ~~~ & ~~~$\rm Exact$~~~ & ~~~$\rm PAA$ ~~~ \\
\hline
  $r^D_{1,0}$ & ~1 & - \\
\hline
  $r^D_{2,0}$ & ~1.84028 & - \\
\hline
  $r^D_{3,0}$ & -3.03222 & [0/1]:~3.38662 \\
\hline
  $r^D_{4,0}$ & -23.2763 & [0/2]:-17.3926 \\
\hline
  $r^D_{5,0}$ & - & [0/3]:-34.1991 \\
\hline
\end{tabular}
\caption{The preferable [0/$n$-1]-type PAA predictions of three-, four-, and five-loop coefficients of $a_{D}|_{\rm PMC}$ under the PMC single-scale approach. The known coefficients (``Exact") are also presented as comparisons.}  \label{pmcpaa1}
\end{table}

\begin{table}[htb]
\begin{tabular}{ccc}
\hline
 ~~~ ~~~ & ~~~$\rm Exact$~~~ & ~~~$\rm PAA$ ~~~ \\
\hline
  $r^{F_3}_{1,0}$ & ~1 & - \\
\hline
  $r^{F_3}_{2,0}$ & ~0.840278 & - \\
\hline
  $r^{F_3}_{3,0}$ & -5.71277 & [0/1]:~0.706067 \\
\hline
  $r^{F_3}_{4,0}$ & -16.0776 & [0/2]:-10.1939 \\
\hline
  $r^{F_3}_{5,0}$ & - & [0/3]:~18.2157 \\
\hline
\end{tabular}
\caption{The preferable [0/$n$-1]-type PAA predictions of three-, four-, and five-loop coefficients of $a_{F_3}|_{\rm PMC}$ under the PMC single-scale approach. The known coefficients (``Exact") are also presented as comparisons. }  \label{pmcpaa2}
\end{table}

Fig.~\ref{RPMC} shows that the two-, three-, four-, and the predicted five-loop predictions are close in shape, especially the predicted five-loop curve almost coincides with the four-loop one. This is due to the fact that the PMC conformal series is free of renormalon divergence~\cite{Beneke:1994qe, Neubert:1994vb, Beneke:1998ui}, which inversely results in a good pQCD convergence. The scheme-independent $D(a_s)|_{\rm{PMC}}$ and $1/C^{\rm GLS}(a_s)|_{\rm PMC}$ have the same conformal coefficients, but as shown by Fig.~(\ref{Qstar}), their PMC scales are not identical, ${\tilde Q}_*\neq{\hat Q}_*$, thus there is a small deviation from unity:
\begin{equation}
D(a_s)|_{\rm PMC}~C^{\rm GLS}(a_s)|_{\rm PMC} \approx 1. \label{eq:GCRPMC0}
\end{equation}
This deviation, though very small in large $Q$ region, e.g. for $Q\simeq10^3$ GeV, the ratio is about $0.998$ for the five-loop prediction, is the intrinsic nature of QCD theory, due to its conformal breaking property.

In GCR (\ref{eq:GCR}), the scheme dependent $\Delta_{\rm csb}$-term has been introduced to collect all the conformal breaking terms. This leads to explicit scheme dependence of the GCR (\ref{eq:GCR}) under conventional scale-setting approach due to the mismatching of $\alpha_s$ and its corresponding expansion coefficients. On the other hand, by applying the PMC single-scale approach, we can achieve an exactly scheme and scale invariant GCR at any fixed order. More explicitly, after applying the PMC single-scale approach, we obtain the following conformal series,
\begin{eqnarray}
a_D(Q)=\sum_{i=1}^n a_{F_3}^i(Q_*), \label{eqGSICR}
\end{eqnarray}
where all the coefficients are exactly equal to $1$, the PMC scale $Q_*$ is independent to the choice of renormalization scale, which can be fixed up to N$^2$LL accuracy by using the known four-loop coefficients\footnote{ The scale $Q_*$ satisfies Eqs.(8-11) of Ref.\cite{Shen:2017pdu}, whose value can be determined by replacing $a_s$ and ${\hat r}_{i,j}$ with $a_{F_3}$ and $r^{D F_3}_{i,j}$. The $r^{D F_3}_{i,j}$ is a function of $r^D_{i,j}$ and $r^{F_3}_{i,j}$~\cite{Brodsky:2013vpa}. One can derive Eqs.(\ref{T0}-\ref{T2}) by substituting those functions into Eqs.(9-11) of Ref.\cite{Shen:2017pdu}. }, i.e.,
\begin{eqnarray}
\ln\frac{Q^2_*}{Q^2}=T_0 +T_1 a_{F_3}(Q) +T_2 a^2_{F_3}(Q),
\label{qstar}
\end{eqnarray}
where
\begin{eqnarray}
T_0&=&r^{F_3}_{2,1}-r^D_{2,1}, \label{T0}\\
T_1&=&2(r^{F_3}_{3,1}-r^D_{3,1}+r^D_{2,0} r^D_{2,1}-r^{F_3}_{2,0} r^{F_3}_{2,1}) \nonumber \\ &&+[r^{F_3}_{3,2}-r^D_{3,2}+(r^D_{2,1})^2-(r^{F_3}_{2,1})^2]\beta_0 ,\\
T_2&=&3(r^{F_3}_{4,1}-r^D_{4,1})-2r^{F_3}_{2,1} r^{F_3}_{3,0}+4r^D_{2,0} r^D_{3,1}+3r^D_{2,1} r^D_{3,0}\nonumber \\
&&-4(r^D_{2,0})^2 r^D_{2,1}-r^D_{2,1} r^{F_3}_{3,0}+(r^{F_3}_{2,0})^2(r^D_{2,1}+5r^{F_3}_{2,1})\nonumber \\
&&-2r^{F_3}_{2,0}(3r^{F_3}_{3,1}-r^D_{3,1}+r^D_{2,0} r^D_{2,1})\nonumber \\
&&+\{3(r^{F_3}_{4,2}-r^D_{4,2})+2r^D_{2,0} r^D_{3,2}+4r^D_{3,1} r^D_{2,1}\nonumber \\
&&-3r^D_{2,0} (r^D_{2,1})^2-2r^{F_3}_{2,1}(3r^{F_3}_{3,1}-r^D_{3,1}+r^D_{2,0} r^D_{2,1})\nonumber \\
&&+r^{F_3}_{2,0}[r^D_{3,2}-3r^{F_3}_{3,2}+6(r^{F_3}_{2,1})^2-(r^D_{2,1})^2]\}\beta_0\nonumber \\
&&+\{r^{F_3}_{4,3}-r^D_{4,3}+2r^D_{3,2}r^D_{2,1}+2(r^{F_3}_{2,1})^3-(r^D_{2,1})^3\nonumber \\
&&+r^{F_3}_{2,1}[r^D_{3,2}-3r^{F_3}_{3,2}-(r^D_{2,1})^2]\}\beta_0^2\nonumber \\
&&+\frac{3}{2}[r^{F_3}_{3,2}-r^D_{3,2}+(r^D_{2,1})^2-(r^{F_3}_{2,1})^2]\beta_1. \label{T2}
\end{eqnarray}
Eq.~(\ref{eqGSICR}) is exactly scheme-independent, which can be treated as a kind of commensurate scale relation (CSR). The CSR has been suggested in Ref.\cite{Brodsky:1994eh} with the purpose of ensuring the scheme-independence of the pQCD approximants among different renormalization schemes, and all the original CSRs suggested in Ref.\cite{Brodsky:1994eh} are at the NLO level. The PMC single-scale approach provides a way to extend the CSR to any orders. A general demonstration on the scheme independence of the CSR (\ref{eqGSICR}) shall be given in next section. As a special case, taking the conformal limit that all $\{\beta_i\}$-terms tend to zero, we have $Q_* \to Q$, and then the relation (\ref{eqGSICR}) turns to the original Crewther relation~\cite{Broadhurst:1993ru}
\begin{eqnarray}
[1+a_D(Q)][1-a_{F_3}(Q)]\equiv 1.
\end{eqnarray}

By applying the PMC single-scale approach, one can obtain similar scheme-independent relations among different observables. The relation (\ref{eqGSICR}) not only provides a fundamental scheme-independent relation but also has phenomenologically useful consequences.

For example, the effective charge $a_{F_3}$ can be related to the effective change $a_R$ of $R$-ratio for the $e^+e^-$ annihilation cross section ($R_{e^+e^-}$). The measurable $R$-ratio can be expressed by the perturbatively calculated Adler function, i.e.,
\begin{eqnarray}
R_{e^+e^-}(s)&=&\frac{1}{2\pi i}\int^{-s+i \epsilon}_{-s-i \epsilon}\frac{D(a_s(Q))}{Q^2}dQ^2 \nonumber\\
&=&3\sum_f q_f^2(1+a_R(Q)),
\end{eqnarray}
where $q_f$ is the electric charge of the active flavor. Similarly, the perturbative series of the effective charge $a_R(Q)$ can be written as
\begin{eqnarray}
a_R&=&r^R_{1,0}a_s + (r^R_{2,0}+\beta_{0}r^R_{2,1})a_s^{2}\nonumber\\
&&+(r^R_{3,0}+\beta_{1}r^R_{2,1}+ 2\beta_{0}r^R_{3,1}+ \beta_{0}^{2}r^R_{3,2})a_s^{3}\nonumber\\
&& +(r^R_{4,0}+\beta_{2}r^R_{2,1}+ 2\beta_{1}r^R_{3,1} + \frac{5}{2}\beta_{1}\beta_{0}r^R_{3,2} \nonumber\\
&&+3\beta_{0}r^R_{4,1}+3\beta_{0}^{2}r^R_{4,2}+\beta_{0}^{3}r^R_{4,3}) a_s^{4}+\mathcal{O}(a^5_s),
\end{eqnarray}
where the coefficients $r^R_{i,j}$ under the $\overline{\rm MS}$-scheme can be derived from Refs.\cite{Baikov:2008jh, Baikov:2010je, Baikov:2012zm, Baikov:2012zn}~\footnote{Here, we will not consider the nonperturbative contributions, which may be important in small $Q^2$-region~\cite{Shifman:1978bx, Jaffe:1982pm, Shuryak:1981kj, Shuryak:1981pi, Ji:1993sv, Braun:1986ty, Balitsky:1989jb, Ross:1993gb, Stein:1994zk, Stein:1995si}, but are negligible for comparatively large $s$ and $Q^2$.}. After applying the PMC single-scale approach, we obtain a relation between $a_R$ and $a_{F_3}$ up to N$^3$LO, i.e.,
\begin{eqnarray}
a_{F_3}(Q)&=& \sum_{i=1}^{n} r_{i,0} a^{i}_R(\bar{Q}_*), \label{aF3}
\end{eqnarray}
whose first four coefficients are
\begin{eqnarray}
r_{1,0}&=&1, \\
r_{2,0}&=&-1, \\
r_{3,0}&=&1+\gamma^{\rm S}_3\left[\frac{99}{8}-\frac{3(\sum_f q_f)^2}{4\sum_f q^2_f}\right],\\
r_{4,0}&=&-1+\frac{(\sum_f q_f)^2}{\sum_f q^2_f}\left(\frac{27\gamma^{\rm NS}_2 \gamma^{\rm S}_3}{16}+\frac{3\gamma^{\rm S}_3}{2}-\frac{3\gamma^{\rm S}_4}{4}\right)\nonumber \\
&&-\gamma^{\rm S}_3 \left(\frac{99}{4}+\frac{891\gamma^{\rm NS}_2}{32}\right) +\frac{99\gamma^{\rm S}_4}{8},
\end{eqnarray}
where $\gamma^{\rm S, NS}_{i}$ are singlet and non-singlet anomalous dimensions which are unrelated to $\alpha_s$-renormalization, and the effective PMC scale $\bar{Q}_*$ can be determined up to N$^2$LL accuracy, which reads
\begin{eqnarray}
\ln\frac{\bar{Q}^2_*}{Q^2}=S_0+S_1 a_R(\bar{Q}_*)+S_2 a_R^2(\bar{Q}_*),
\label{qstarF3}
\end{eqnarray}
where
\begin{eqnarray}
S_0=&&-\frac{r_{2,1}}{r_{1,0}}, \\
S_1=&&\frac{ \beta _0 (r_{2,1}^2-r_{1,0} r_{3,2})}{r_{1,0}^2}+\frac{2 (r_{2,0} r_{2,1}-r_{1,0} r_{3,1})}{r_{1,0}^2},
\end{eqnarray}
and
\begin{eqnarray}
S_2=&&\frac{3 \beta _1 (r_{2,1}^2-r_{1,0}r_{3,2})}{2 r_{1,0}^2}\nonumber\\
&&\!\!\!\!\!\!\!\!  +\frac{4(r_{1,0} r_{2,0} r_{3,1}-r_{2,0}^2 r_{2,1})+3(r_{1,0} r_{2,1} r_{3,0}-r_{1,0}^2 r_{4,1})}{r_{1,0}^3} \nonumber \\
&&\!\!\!\!\!\!\!\!  +\frac{ \beta _0  (6 r_{2,1} r_{3,1} r_{1,0}-3 r_{4,2} r_{1,0}^2+2 r_{2,0} r_{3,2} r_{1,0}-5 r_{2,0} r_{2,1}^2)}{r_{1,0}^3}\nonumber\\
&&\!\!\!\!\!\!\!\!  +\frac{ \beta _0^2 (3 r_{1,0} r_{3,2} r_{2,1}- 2r_{2,1}^3- r_{1,0}^2 r_{4,3})}{ r_{1,0}^3}.
\end{eqnarray}

Experimentally, the effective charge $a_R$ has been constrained by measuring $R_{e^+e^-}$ above the thresholds for the production of $(c\bar{c})$-bound state~\cite{Mattingly:1993ej}, i.e.
\begin{equation}
a^{\rm exp}_R(\sqrt{s}=5 {\rm GeV})\simeq 0.08 \pm0.03. \label{ar}
\end{equation}
Substituting it into Eq.~(\ref{aF3}), we obtain
\begin{eqnarray}
a_{F_3}(Q=12.58^{+1.48}_{-1.26}~\rm GeV)
&=&0.073^{+0.025}_{-0.026}. \label{aF3result1}
\end{eqnarray}
which is consistent with the PMC prediction derived directly from Eq.~(\ref{CGLSpmc}) within errors, e.g.
\begin{equation}
a_{F_3}(Q=12.58^{+1.48}_{-1.26}~{\rm GeV})|_{\rm PMC}=0.063^{+0.002+0.001}_{-0.001-0.001},
\end{equation}
where the first error is for $\Delta Q=\left(^{+1.48}_{-1.26}\right)~{\rm GeV}$ and the second error is for $\Delta\alpha_{s}(M_Z) =0.1179\pm0.0011$~\cite{Zyla:2020zbs}. At present, the GLS sum rules is measured at small $Q^2$-values~\cite{Leung:1992yx, Quintas:1992yv}, an extrapolation of the data gives~\cite{Kataev:1994rj}
\begin{equation}
a^{\rm ext}_{F_3}(Q=12.25~{\rm GeV}) \simeq 0.093\pm 0.042,
\end{equation}
which agrees with our prediction within errors.

\section{A demonstration of the scheme independence of commensurate scale relation}

In this section, we give a novel demonstration of the scheme independence of CSR to all orders by relating different pQCD approximants within the effective charge method. The effective charge $a_A$ can be expressed as a perturbative series over another effective charge $a_B$,
\begin{eqnarray}
a_A &=& r^{\rm AB}_{1,0}a_{B} + (r^{\rm AB}_{2,0}+\beta_{0}r^{\rm AB}_{2,1})a_{B}^{2}+ \nonumber\\
&& (r^{\rm AB}_{3,0}+\beta_{1}r^{\rm AB}_{2,1}+ 2\beta_{0}r^{\rm AB}_{3,1}+ \beta_{0}^{2}r^{\rm AB}_{3,2})a_{B}^{3}+ \nonumber\\
&& (r^{\rm AB}_{4,0}+\beta^{B}_{2}r^{\rm AB}_{2,1}+ 2\beta_{1}r^{\rm AB}_{3,1} + \frac{5}{2}\beta_{1}\beta_{0}r^{\rm AB}_{3,2}+ \nonumber\\
&& 3\beta_{0}r^{\rm AB}_{4,1}+3\beta_{0}^{2}r^{\rm AB}_{4,2}+\beta_{0}^{3}r^{\rm AB}_{4,3}) a_{B}^{4}+\mathcal{O}(a_{B}^5), \label{eq:aAaB}
\end{eqnarray}
where $a_A$ and $a_B$ stand for the effective charges under arbitrary schemes $A$ and $B$, respectively. The previously mentioned $a_D$ and $a_{F_3}$ are such kind of effective charges. The $\{\beta_i\}$-functions are usually calculated under the $\overline{\rm MS}$-scheme, and the $\{\beta_i\}$-functions for $A/B$ scheme can be obtained by using its relation to the $\overline{\rm MS}$-scheme one, i.e. $\beta^{A/B}={\partial \alpha_{s, A/B}}/{\partial \alpha_{s, \overline{\rm MS}}}{\beta^{\overline{\rm MS}}}$. The effective charge $a_B$ at any scale $\mu$ can be expanded in terms of a $C$-scheme coupling $\hat{a}_B(\mu)$ at the same scale~\cite{Wu:2018cmb}, i.e.,
\begin{eqnarray}
a_{B} &=& {\hat a}_{B}+C\beta_0 {\hat a}_{B}^2+ \left(\frac{\beta_2^B}{\beta_0}- \frac{\beta_1^2}{\beta_0^2}+\beta_0^2 C^2+\beta_1 C\right) {\hat a}_{B}^3 \nonumber\\
& & +\left[\frac{\beta_3^B}{2 \beta _0}-\frac{\beta_1^3}{2 \beta _0^3}+\left(3 \beta_2^B-\frac{2 \beta _1^2}{\beta _0}\right) C+\frac{5}{2}\beta_0 \beta_1 C^2 \right. \nonumber\\
&& +\beta_0^3 C^3  \Big] {\hat a}_{B}^4 + {\cal O}({\hat a}_{B}^5),
\label{eq:Expandhata}
\end{eqnarray}
where by choosing a suitable $C$, the coupling ${\hat a}_{B}$ can be equivalent to $a_B$ defined for any scheme at the same scale, i.e. $a_B={\hat a}_{B}|_{C}$. By using the $C$-scheme coupling, the relation (\ref{eq:aAaB}) becomes
\begin{eqnarray}
a_A&=&r_1 {\hat a}_{B} + (r_2 + \beta_0 r_1 C) {\hat a}_{B}^{2}+\bigg[r_3 + \bigg(\beta_1 r_1+2\beta_0 r_2\bigg)C\nonumber\\
&&+\beta_0^2 r_1 C^2+ r_1\bigg(\frac{\beta_2^B}{\beta_0}-\frac{\beta_1^2}{\beta_0^2}\bigg)\bigg] {\hat a}_{B}^{3} \nonumber\\
&&+\bigg[r_4 + \bigg(3\beta_0 r_3+2\beta_1 r_2+3 \beta_2^B r_1-\frac{2\beta_1^2 r_1}{\beta_0}\bigg)C \nonumber\\
&&+\bigg(3\beta_0^2 r_2+\frac{5}{2} \beta_1 \beta_0 r_1\bigg)C^2+r_1\beta_0^3 C^3 \nonumber \\
&&+r_1\bigg(\frac{\beta_3^B}{2\beta_0}-\frac{\beta_1^3}{2\beta_0^3}\bigg)
 +r_2\bigg(\frac{2\beta_2^B}{\beta_0}-\frac{2\beta_1^2}{\beta_0^2}\bigg)\bigg] {\hat a}_{B}^{4}\nonumber \\
&&+\mathcal{O}({\hat a}_{B}^5),
\end{eqnarray}
where the coefficients $r_i$ are
\begin{eqnarray}
r_1 &=& r^{\rm AB}_{1,0}, \label{r1-conf-beta}\\
r_2 &=& r^{\rm AB}_{2,0}+\beta_{0}r^{\rm AB}_{2,1}, \label{r2-conf-beta}\\
r_3 &=& r^{\rm AB}_{3,0}+\beta_{1}r^{\rm AB}_{2,1}+ 2\beta_{0}r^{\rm AB}_{3,1}+ \beta_{0}^{2}r^{\rm AB}_{3,2}, \label{r3-conf-beta}\\
r_4 &=& r^{\rm AB}_{4,0}+\beta^{B}_{2}r^{\rm AB}_{2,1}+ 2\beta_{1}r^{\rm AB}_{3,1} + \frac{5}{2}\beta_{1}\beta_{0}r^{\rm AB}_{3,2} \nonumber\\
&&+3\beta_{0}r^{\rm AB}_{4,1}+3\beta_{0}^{2}r^{\rm AB}_{4,2}+\beta_{0}^{3}r^{\rm AB}_{4,3}. \label{r4-conf-beta}
\end{eqnarray}
Following the standard PMC single-scale approach, we obtain the following CSR
\begin{equation}
a_A(Q)=\sum_{i=1}^n r^{\rm AB}_{i,0}{\hat a}_{B}^i(Q_{**}),  \label{ABCSR}
\end{equation}
where the effective PMC scale $Q_{**}$ is obtained by vanishing all nonconformal terms, which can be expanded as a power series over ${\hat a}_{B}(Q_{**})$, i.e.,
\begin{eqnarray}
\ln\frac{Q_{**}^2}{Q^2} = \sum^{n-2}_{i=0} \hat{S}_i {\hat a}_{B}^i(Q_{**}),
\label{eq:C-PMC-scale}
\end{eqnarray}
whose first three coefficients are
\begin{eqnarray}
\hat{S}_0 &=& -\frac{r^{\rm AB}_{2,1}}{r^{\rm AB}_{1,0}}-C, \label{S0}\\
\hat{S}_1 &=& \frac{2\left(r^{\rm AB}_{2,0}r^{\rm AB}_{2,1}-r^{\rm AB}_{1,0}r^{\rm AB}_{3,1}\right)}{(r^{\rm AB}_{1,0})^2}
+\frac{(r^{\rm AB}_{2,1})^2-r^{\rm AB}_{1,0} r^{\rm AB}_{3,2}}{(r^{\rm AB}_{1,0})^2}\beta_0 \nonumber\\&&+\frac{\beta_1^2}{\beta_0^3}-\frac{\beta^{B}_2}{\beta_0^2}, \label{S1}
\end{eqnarray}
and
\begin{eqnarray}
\hat{S}_2 &=&\frac{3r^{\rm AB}_{1,0}r^{\rm AB}_{2,1}r^{\rm AB}_{3,2} -(r^{\rm AB}_{1,0})^2r^{\rm AB}_{4,3} -2(r^{\rm AB}_{2,1})^3}{(r^{\rm AB}_{1,0})^3}\beta_0^2 + \nonumber\\
&& \hspace{-0.8cm}\frac{3r^{\rm AB}_{1,0}\left(2r^{\rm AB}_{2,1} r^{\rm AB}_{3,1} -r^{\rm AB}_{1,0} r^{\rm AB}_{4,2}\right)+ r^{\rm AB}_{2,0}\left[2 r^{\rm AB}_{1,0} r^{\rm AB}_{3,2}-5 (r^{\rm AB}_{2,1})^2\right]}{(r^{\rm AB}_{1,0})^3}\beta_0 \nonumber\\
&&\hspace{-0.8cm}+\frac{3r^{\rm AB}_{1,0}\left(r^{\rm AB}_{3,0}r^{\rm AB}_{2,1} -r^{\rm AB}_{1,0}r^{\rm AB}_{4,1}\right)+4r^{\rm AB}_{2,0} \left(r^{\rm AB}_{1,0}r^{\rm AB}_{3,1}-r^{\rm AB}_{2,0} r^{\rm AB}_{2,1}\right)}{(r^{\rm AB}_{1,0})^3} \nonumber\\
&&\hspace{-0.8cm}+ \frac{3\left[(r^{\rm AB}_{2,1})^2-r^{\rm AB}_{1,0}r^{\rm AB}_{3,2}\right]}{2(r^{\rm AB}_{1,0})^2}\beta_1- \frac{\beta_1^3}{2\beta_0^4}+\frac{\beta^{B}_2 \beta_1}{\beta_0^3}-\frac{\beta^{B}_3}{2\beta_0^2}.\label{S2}
\end{eqnarray}
One may observe that only the LL coefficient $\hat{S}_{0}$ depends on the scheme parameter $C$, and all the higher order coefficients $\hat{S}_{i}$ ($i\geq1$) are free of $C$.

Moreover, the $C$-scheme coupling ${\hat a}_{B}(Q_*)$ satisfies the following relation~\cite{Wu:2018cmb},
\begin{equation}
\frac{1}{{\hat a}_{B}(Q_{**})}+\frac{\beta_1}{\beta_0} \ln {\hat a}_{B}(Q_{**}) = \beta_0\left(\ln\frac{Q_{**}^2}{\Lambda^2}+C\right).
\end{equation}
Substituting Eq.~(\ref{eq:C-PMC-scale}) into the right hand side of the equation, we obtain
\begin{eqnarray}
&& \frac{1}{{\hat a}_{B}(Q_{**})}+\frac{\beta_1}{\beta_0} \ln {\hat a}_{B}(Q_{**}) \nonumber\\
&=& \beta_0 \bigg[ \ln\frac{Q^2}{\Lambda^2}-\frac{r^{\rm AB}_{2,1}}{r^{\rm AB}_{1,0}}+\sum^{n-2}_{i= 1} \hat{S}_i  {\hat a}_{B}^i(Q_{**}) \bigg]. \label{Csas2}
\end{eqnarray}
By using this equation, we can derive a solution for ${\hat a}_{B}(Q_{**})$. Because all the coefficients in  Eq.~(\ref{Csas2}) are free of $C$ at any fixed order, the magnitude of ${\hat a}_{B}(Q_{**})$ shall be exactly free of $C$. Together with the scheme-independent conformal coefficients and the fact that the value of $C$ can be chosen to match any renormalization scheme, we can conclude that the CSR (\ref{ABCSR}) is exactly scheme independent. This demonstration can be extended to all orders.

\section{Summary}

The PMC provides a systematic approach to determine an effective $\alpha_s$ for a fixed-order pQCD approximant. By using the PMC single-scale approach, the determined effective $\alpha_s$ is scale-invariant, which is free of any choice of renormalization scale. Because all nonconformal terms have been eliminated, the resultant pQCD series shall be scheme independent, well satisfying the requirements of RGI. Furthermore, by applying the PMC single-scale approach, we obtain a scheme-independent GCR, $D(a_s)|_{\rm PMC}C^{\rm GLS}(a_s)|_{\rm PMC} \approx 1$, which provides a significant connection between the Adler function and the GLS sum rules. We have shown that their corresponding effective couplings satisfy a scheme-independent CSR, $a_D(Q)=\sum_{i=1}^{n} a_{F_3}^i(Q_*)$. Furthermore, we obtain a CSR that relates the effective charge $a_{F_3}$ to the effective charge of $R$-ratio, $a_{F_3}(Q)=\sum_{i=1}^{4} r_{i,0} a^{i}_R(\bar{Q}_*)$. This leads to $a_{F_3}(Q=12.58^{+1.48}_{-1.26}~\rm GeV) = 0.073^{+0.025}_{-0.026}$, which agrees with the extrapolated measured value within errors. A demonstration on the scheme-independence of the CSR has been presented. The scheme- and scale- independent CSRs shall provide important tests of pQCD theory.

\hspace{1cm}

{\bf Acknowledgements:} This work is partly supported by the Chongqing Graduate Research and Innovation Foundation under Grant No.ydstd1912 and No.CYB21045, the National Natural Science Foundation of China under Grant No.11625520 and No.12047564, and the Fundamental Research Funds for the Central Universities under Grant No.2020CQJQY-Z003.

\end{document}